\documentclass[twocolumn,showpacs,preprintnumbers,amsmath,amssymb,superscriptaddress,aps,prc]{revtex4-1}
\usepackage{graphicx}
\usepackage{xcolor}
\usepackage{subfigure}
\def\scr{\rm\scriptscriptstyle }
\usepackage{amssymb}
\usepackage{graphicx}
\usepackage{subfigure}
\usepackage{amsmath}
\usepackage[mathscr]{eucal}

\def\scr{\rm\scriptscriptstyle }
\def\HPP{\mathbb{H}_{\rm\scriptscriptstyle PP}}
\def\HPQ{\mathbb{H}_{\rm\scriptscriptstyle PQ}}
\def\HQP{\mathbb{H}_{\rm\scriptscriptstyle QP}}
\def\HQQ{\mathbb{H}_{\rm\scriptscriptstyle QQ}}

\def\VPP{\mathbb{V}_{\rm\scriptscriptstyle PP}}

\def\PsiP{\Psi^{\rm\scriptscriptstyle(+)}_{\rm\scriptscriptstyle P}}
\def\PsiQ{\Psi_{\rm\scriptscriptstyle Q}}

\begin{document}

\title{	Theoretical considerations about heavy-ion fusion in potential scattering}

\author{L.F. Canto}
	\email{canto@if.ufrj.br}
	\affiliation{Instituto de F\'{\i}sica, Universidade Federal do Rio de Janeiro, CP 68528, 21941-972, Rio de Janeiro, RJ, Brazil}
	\affiliation{Instituto de F\'{\i}sica, Universidade Federal Fluminense, Av. Litoranea s/n, Gragoat\'{a}, Niter\'{o}i, R.J., 24210-340, Brazil}
\author{R. Donangelo}
	\email{donangel@fing.edu.uy}
	\affiliation{Instituto de F\'{\i}sica, Facultad de Ingenier\'{\i}a, Universidad de la Rep\'ublica, C.C. 30, 11000 Montevideo, Uruguay}

\author{M. S. Hussein}
	\email{hussein@if.usp.br}
	\affiliation{Instituto de F\'isica, Universidade de S\~ao Paulo, C.P. 66318, 05314-970, S\~ao Paulo, SP, Brazil}
	\affiliation{Instituto de Estudos Avan\c{c}ados, Universidade de S\~ao Paulo, C.P. 72012, 05508-970, 
		S\~ao Paulo, SP, Brazil}
	\affiliation{Departamento de F\'{i}sica, Instituto Tecnol\'{o}gico de Aeron\'{a}utica, CTA, S\~{a}o Jos\'{e} dos Campos, S\~ao Paulo, SP, Brazil}
\author{P. Lotti} 
	\email{paolo.lotti@pd.infn.it}
	\affiliation{INFN, Sezione di Padova, Via F. Marzolo 8, 35131, Padova, Italy }
\author{J. Lubian}
	\email{lubian@if.uff.br}
	\affiliation{Instituto de F\'{\i}sica, Universidade Federal Fluminense, Av. Litoranea s/n, Gragoat\'{a}, Niter\'{o}i, R.J., 24210-340, Brazil}

\author{J. Rangel}
	\email{jeannie@if.uff.br}
	\affiliation{Instituto de F\'{\i}sica, Universidade Federal Fluminense, Av. Litoranea s/n, Gragoat\'{a}, Niter\'{o}i, R.J., 24210-340, Brazil}

\begin{abstract}
We  carefully compare the one-dimensional WKB barrier tunneling  model,  and the one-channel 
Sch\"odinger equation with a complex optical potential calculation of heavy-ion fusion, for a light 
and a heavy system. 
It is found that the major difference between the two approaches occurs around the critical energy, 
above which the effective potential for the grazing angular momentum ceases to exhibit a pocket. 
The value of this critical energy is shown to be strongly dependent on the nuclear potential at short 
distances, on the inside region of the Coulomb barrier, and this dependence is much more important 
for heavy systems. 
Therefore the nuclear fusion process is expected to provide information on the nuclear potential in 
this inner region. 
We compare calculations with available data to show that the results are consistent with this expectation.

\end{abstract}

\maketitle

\section{Introduction}

Nuclear collisions involve several degrees of freedom. Besides the projectile-target separation vector, 
${\bf r}$, the collision depends on intrinsic coordinates of the collision partners, which are coupled with 
${\bf r}$ by Coulomb and Nuclear forces~\cite{CaH13,Sat83}. 
In this way, the collision may lead to various final states of the systems (channels). 
In addition to elastic scattering, it may undergo direct reactions, like inelastic scattering, transfer and breakup, or 
fuse to form a compound nucleus (CN). 
The simplest quantum mechanical treatment of a nuclear collision, referred to as {\it potential scattering}, 
ignores all intrinsic degrees of freedom. 
It approximates the problem by a collision of two point particles, interacting through a real potential, $V({\bf r})$.  
Clearly, this approach can only make predictions for elastic scattering. However, it is necessary to take into 
account the attenuation of the incident wave, resulting from transitions to non-elastic channels. 
Owing to these transitions, the current associated with the elastic wave function does not satisfy the continuity 
equation. 
This would be inconsistent with the wave function of a hermitian Hamiltonian. 
To fix this problem, one simulates the effects of non-elastic channels by the inclusion of a negative imaginary 
part in the potential.\\

However, potential scattering is a very poor treatment of nucleus-nucleus collisions. 
More satisfactory results can be obtained through the Coupled-Channel (CC) approach. 
In this method, the wave function is expanded over a set of intrinsic states of the system and the expansion 
is inserted into the  Schr\"odinger equation with the full Hamiltonian. 
In this way, one gets a set of coupled equations for the wave functions in these channels.  
If the expansion contained all relevant channels, one would obtain realistic predictions for the experimental 
cross sections. 
However, this condition is not satisfied in heavy-ion collisions, owing to the fusion channel. 
The formation of a compound nucleus and its subsequent decay are very complicated processes, that cannot 
be handled in the coupled channel approach. 
Nevertheless, fusion and its influence on direct reactions must be taken into account. 
They can be estimated through the inclusion of an imaginary potential in the Hamiltonian. 
This potential must be negative and very strong, to absorb completely the current that reaches the inner region 
of the Coulomb barrier. 
It is believed that the details of this potential are not important, provided that it produces strong short-range 
absorption.\\

An alternative way to handle fusion, is to keep the potential real and solve the radial equations with ingoing 
wave boundary conditions (IWBC) at some radial distance $r=R_{\rm in}$, located in the inner region of the barrier. 
This procedure is adopted by some authors, and used in the CCFULL~\cite{HRK99} computer code. 
The IWBC assumes that there are no reflected waves at $R_{\rm in}$, which implies that the incident wave is 
completely absorbed at $r <R_{\rm in}$. 
Thus, it is expected to be equivalent to solving the radial equation with the usual boundary conditions at $r=0$, 
but with a complex potential which produces total absorption in this region, and is not active elsewhere.\\

The purpose of the present work is to investigate the dependence of the fusion cross section on reasonable 
choices of the interaction.
For simplicity, our study is restricted to potential scattering with real nuclear potentials evaluated by some 
version of the folding model. 
These potentials take into account the nuclear densities but ignores the nuclear structure properties of the 
collision partners. 
It is well known that these properties may strongly affect sub-barrier fusion. 
Therefore, we consider only collisions at above-barrier energies, where the fusion cross section predicted 
by potential scattering and the ones obtained in coupled channel calculations are similar. 
Although this may not happen in fusion reactions with weakly bound projectiles, it does not affect our 
conclusions, since collisions of this kind are not considered here. \\

The paper is organized as follows. 
In Sec. II, we discuss the basic aspects of fusion in potential scattering. 
In Sec. III we investigate the influence of different commonly used treatments of absorption and choices of 
the nuclear potential on the fusion cross section,  considering as examples the cases of a heavy and a light system. 
Finally, in Sec. IV we present the conclusions of our work.\\ 

\section{The fusion cross section in potential scattering}

The collision dynamics in potential scattering is governed by the Hamiltonian,
\begin{equation}
H = K+V(r),
\label{Hpotscat}
\end{equation}
where $K$ is the kinetic energy operator associated with the relative motion of the collision partners and $V(r)$ 
is the total interaction between them. 
Usually it is written as, 
\begin{equation}
V(r) = U(r) \,+\,i\,W(r).
\label{UW}
\end{equation}
The real part of the interaction can be written,
\begin{equation}
U(r) = U_{\scr C}(r) \,+\,U_{\scr N}(r) ,
\label{V-real}
\end{equation}
where $U_{\scr C}(r)$ and $U_{\scr N}(r)$ are respectively the Coulomb and the nuclear potentials, 
and $i\,W(r)$ is a negative imaginary function that renders the Hamiltonian non-hermitian. 
In this way, the absorption of the incident wave resulting from excitations of non-elastic channels is accounted for. \\

The scattering wave function for a collision with energy $E$ satisfies the Schr\"odinger equation
\begin{equation}
\big[E-H \big]\,\psi^{\scr (+)}({\bf r}) = 0 .
\label{Sch}
\end{equation}
To determine this wave function, one carries out partial-waves expansions in the above equations. 
In this way, one gets one equation for each radial wave function, $u_l(k,r)$ (for simplicity, we are 
neglecting spins), involving the effective $l$-dependent potential
\begin{equation}
U_l(r) = U_{\scr C}(r) + U_{\scr N}(r) + \frac{\hbar^2\,l(l+1)}{2\mu r^2}.
\label{Ul}
\end{equation} 
These equations should be solved to build solutions with the scattering boundary conditions,
\begin{equation}
u_l(k,r \rightarrow \infty) = \frac{i}{2} \big[ H^{\scr (-)}_l(kr,\eta)\, -\, S_{{\scr N},{\sc l}}(E)\,H^{\scr (+)}_l(kr,\eta) \big],
\label{uasymp}
\end{equation}
where  $k$ is the wave number, $\eta$ is the Sommerfeld parameter and $H^{\scr (-)}_l(kr,\eta)$ ($H^{\scr (+)}_l(kr,\eta)$) 
is the Coulomb wave function with ingoing (outgoing) behavior. 
Above, $S_{{\scr N},{\sc l}}(E)$ is the $l^{\rm th}$ component of the nuclear $S$-matrix.\\

The potential of Eq.~(\ref{Ul}) is a combination of the attractive nuclear potential with the repulsive 
Coulomb and centrifugal potentials.
For low partial-waves, $U_l(r)$ has a maximum at $r=R_l$ and a pocket at a smaller distance. 
As the angular momentum increases, the centrifugal potential becomes more important and above 
a certain value, which we denote by $l_{\rm cr}$, the pocket disappears. 
Then,  $U_l(r)$ becomes repulsive for any value of $r$.\\

The fusion cross section is given by the expression 
\begin{equation}
\sigma_{\scr F}(E) = \frac{\pi}{k^2}\ \sum_{l=0}^\infty\ (2l+1)\, \mathcal{P}_{\scr F}(l,E)  ,
\label{fusion1}
\end{equation}
where $\mathcal{P}_{\scr F}(l,E) $ stands for the fusion probability at the $l^{\rm th}$ partial-wave 
in a collision with energy $E$. 
This probability is given by the product
\begin{equation}
\mathcal{P}_{\scr F}(l,E) = \mathcal{P}_{\rm abs}(l,E) \times \mathcal{P}_{\scr CN}(l,E).
\label{Pfus}
\end{equation}
The factor  $\mathcal{P}_{\rm abs}(l,E)$ is the probability that the incident wave does not emerge 
in the elastic channel. 
It is measured by the amount of violation of unitarity of the $S$-matrix, through the expression
\begin{equation}
\mathcal{P}_{\rm abs}(l,E) = 1 - \left| S_{{\scr N},{\sc l}}(E) \right|^2 .
\label{Pabs}
\end{equation}
Of course, the $S$-matrix would be unitary in a many-body description of the collision where all 
channels were explicitly taken into account. 
This is not the case in potential scattering. 
In this simplified approach, the loss of flux is usually simulated by an imaginary potential.  
The second factor in Eq.~(\ref{Pfus}), $\mathcal{P}_{\scr CN}(l,E)$, is the probability that the doorway 
state associated with the absorption evolves to CN formation.

\medskip

Alternatively, one can keep the potential real and simulate the loss of flux by an IWBC. 
The calculation of the $S$-matrix is performed through the following steps:
First, one integrates the radial equation numerically, from an internal point to a limiting radial distance, 
$\bar{R}$, where the interaction reduces to the Coulomb potential of two point charges. 
In calculations with a complex potential  the internal point is the origin. 
In the IWBC approach it is some point in the inner region of the barrier, usually the minimum of the 
pocket of the $l$-dependent potential. 
The radial wave function is then assumed to behave as an ingoing wave at this point, and the initial 
conditions are determined within the WKB approximation. 
In both cases, the S-matrix is determined matching the numerically evaluated logarithmic derivative 
at $\bar{R}$ with the one obtained from the asymptotic expression of Eq.~(\ref{uasymp}). 
An important feature of the IWBC approach is that the absorption probability vanishes at angular 
momenta higher than its critical value, $l_{\rm cr}$, above which there is no pocket in the potential. \\


\subsection{The optical potential}


Some experimental analyses of scattering data are based on potential scattering theory. 
Frequently, they determine the complex nuclear potential, referred to as the  {\it optical potential}, 
by fitting elastic cross sections. 
On the other hand, the optical potential can be formally derived from many-body scattering 
theory. In this approach, the dynamics is governed by the full Hamiltonian of the system (blackboard bold 
fonts denote operators acting on both collision and intrinsic degrees of freedom),
\begin{equation}
\mathbb{H}= K + h + \mathbb{V}.
\label{TotH}
\end{equation}
Above, $K$ is the kinetic energy operator of Eq.~(\ref{Hpotscat}), $h$ is the intrinsic Hamiltonian of the 
system, and $\mathbb{V}$ is the operator representing the interaction between the projectile and the target. 
It depends both on the collision coordinate, ${\bf r}$, and the intrinsic degrees of freedom. 
The full scattering state is then the solution of the Schr\"odinger equation
\begin{equation}
\big[E - \mathbb{H}  \big]\ \big| \Psi^{\scr (+)} \big\rangle = 0,
\label{many-body eq}
\end{equation}
with scattering boundary conditions. \\

The derivation of the optical potential can be made formally but quite  transparently within the 
Feshbach theory. 
Denoting by $P$ the elastic channel projection operator and by $Q$ the projector projector on all other 
channels, both closed and open, one can decompose Eq.~(\ref{many-body eq}) into the coupled equations
\begin{eqnarray}
\big[ E - \HPP \big] \, \big| \PsiP \big\rangle &=&\HPQ\, \big| \PsiQ \big\rangle \label{EqPsiP} \\
\big[ E - \HQQ \big] \, \big| \PsiQ \big\rangle &=& \HQP\, \big| \PsiQ \big\rangle \label{EqPsiQ}.
\end{eqnarray}
Above, we adopted the short-range notations:
$\mathbb{H}_{\scr AB} = A\mathbb{H}B$, where $A$ and $B$ stand for any of the two projetors,  
$\Psi^{\scr (+)}_{\scr P} = P \Psi^{\scr (+)}$ and $\Psi_{\scr Q} = Q \Psi^{\scr (+)}$.
At this stage, we assume that only closed channels (compound nucleus) are coupled to the elastic channel. \\

Solving Eq.~(\ref{EqPsiQ}) for $\Psi_{\scr Q}$ and inserting the result into Eq.~(\ref{EqPsiP}),  we obtain 
the effective equation for the elastic component of the wave function,
\begin{equation}
\left[ E - K - \VPP - \mathbb{V}_{\rm eff} \right] \big| \PsiP \big\rangle = 0,
\label{effEq}
\end{equation}
where $\VPP = P\mathbb{V}P$.  This equation involves the potential constrained to the sub-space 
of the elastic channel, $\VPP$, plus the additional term
\begin{equation}
\mathbb{V}_{\rm eff} = \HPQ\,\frac{1}{E - \HQQ}\,\HQP.
\label{Veff}
\end{equation}

The projectors can be expressed in terms of the eigenfunctions of $h$. Denoting them by 
$\left| \varphi_\alpha\right)$, with $\left| \varphi_0\right)$ standing for the ground state, they are given by
\begin{equation}
P = \left| \varphi_0 \right)\, \left(\varphi_0\right| ;\qquad Q = \sum_{\alpha\ne 0} \left| \varphi_\alpha \right)\, \left(\varphi_\alpha\right| .
\label{projectors}
\end{equation}
Inserting the above equations into Eq.~(\ref{effEq}), one gets the Schr\"odinger equation for potential scattering,
\begin{equation}
\left[ E - K - \bar{U} -V_{\rm eff} \right] \big| \psi^{\scr (+)} \big\rangle = 0,
\label{effEq-1}
\end{equation}
where, $\big|\psi^{\scr (+)} \big\rangle$ is the scattering state in the space of the collision degrees of freedom. 
Above,
\begin{equation}
\bar{U} = \big(\varphi_0\big|\, \mathbb{V}\,\big| \varphi_0 \big).
\label{UPP}
\end{equation}
and
\begin{equation}
V_{\rm eff} = \Big(\varphi_0\Big|\, \HPQ\,\frac{1}{E - \HQQ}\,\HQP\,\Big| \varphi_0 \Big)
\label{Veff-1}
\end{equation}
are operators acting exclusively on the ${\bf r}$-space.


\subsubsection{The bare potential}


The potential $U$ of Eq.~(\ref{UPP}) is clearly real. 
It represents the interaction between the collision partners when they are in their ground states,
that is, when couplings to non-elastic channels are completely ignored. 
It can be written as
\begin{equation}
\bar{U}(r) = U_{\scr C}(r) + \bar{U}_{\scr N}(r),
\label{UcUn}
\end{equation}
where $U_{\scr C}$ and $\bar{U}_{\scr N}$ are respectively the Coulomb and nuclear components of $\bar{U}$.

\bigskip

Adopting a microscopic point of view, where the intrinsic degrees of freedom are the nucleon coordinates, 
the bare potential in the coordinate representation can be evaluated by the folding model. 
Then, neglecting nucleon exchange, one gets 
\begin{equation}
\bar{U}_{\scr N}({\bf r})=\int d{\bf r}^\prime\,d{\bf r}^{\prime\prime}\ \rho_{\rm\scriptscriptstyle P}({\bf r}^\prime)\,
v({\bf r}-{\bf r}^{\prime}+{\bf r}^{\prime\prime}) \,\rho_{\rm\scriptscriptstyle T}({\bf r}^{\prime\prime}).
\label{dfold}
\end{equation}
Above, $\rho_{\scr P}$ and $\rho_{\scr T}$ are respectively the densities of the projectile and the target, 
and $v$ is a conveniently chosen nucleon-nucleon interaction. 
A frequent trend in the literature is to adopt Michigan's M3Y nucleon-nucleon interaction~\cite{BBM77}. 
We consider two versions of the folding potential: The S\~ao Paulo potential (SPP)~\cite{CCP78,CPH97,CCG02} 
and the Aky\"uz-Winther potential (AW)~\cite{BrW04,AkW81}. 
The SPP has two advantages. 
The first is that it restores, in an approximate way, exchange effects neglected in the folding integral. 
The second is that the authors developed a computer code to evaluate the integral of Eq.~(\ref{dfold}) using 
the most realistic densities available in the literature. 
On the other hand, there is the disadvantage that this code has not been published, and therefore it is 
not widely available. 
The AW potential has the disadvantage of being less accurate. 
It was developed in three steps. 
First, the authors used approximate analytical expressions for the densities, to simplify the folding integral. 
Second, they evaluated the potential for a large number of systems in different mass ranges, and fitted the 
potentials by WS functions. 
The fits aimed at reproducing the potential in the barrier region. 
Finally, they obtained approximate analytical expressions for the WS parameters, in terms of the mass 
numbers of the collision partners. 
In this way, the evaluation of the WS potential is extremely simple. 
Nevertheless their different origins, the barriers of the SPP and the AW potential are quite similar. 
This point will be discussed further in section \ref{AWvsSPP}.


\subsubsection{The effective coupling potential}


Now we consider the effective potential of Eq.~(\ref{Veff-1}), which accounts for the influence of CN 
couplings on the elastic wave function. 
The most important consequence of these couplings is the partial absorption of the incident wave, 
associated with the populations of CN states. 
Surprisingly, the potential of Eq.~(\ref{Veff}) is real. 
Furthermore, this potential has poles at $E = \HQQ$. 
This very strong energy dependence of the effective potential renders Eq.~(\ref{Veff-1}) useless. \\

The above mentioned shortcomings can be eliminated through energy averaging. 
One chooses an interval in energy, $I$,  which encompasses many compound nucleus resonances. 
This procedure leads to the complex potential~\cite{LeF73,CaH13},
\begin{equation}
V_{\rm eff} = \Big( \varphi_0 \Big|\, \HPQ\,\frac{1}{E - \HQQ + i I/2}\,\HQP\,\Big| \varphi_0 \Big),
\label{Veff-2}
\end{equation}
This potential can be written as 
\begin{equation}
\mathbb{V}_{\rm eff} = \Delta U + i\,W.
\label{complrex Veff}
\end{equation}
The real part of $\mathbb{V}_{\rm eff}$,
\begin{equation}
 \Delta U = \Big(\varphi_0 \Big|\, \HPQ\, \mathcal{P}\left\{ \frac{1}{E - \HQQ + i I/2}\right\}\,\HQP\,\Big|\varphi_0 \Big)
\label{Delta U}
\end{equation}
with $ \mathcal{P}$ standing for the principal value, is a small correction to the potential $U$ of Eq.~(\ref{UPP}). 
It is usually neglected. 
On the other hand, the imaginary part of $\mathbb{V}_{\rm eff}$,
\begin{equation}
W =   - \pi\ \Big(\varphi_0 \Big|\, \HPQ\ \left[\frac{I/2}{\left( E - \HQQ \right)^{2} + I^{2}/4}\right]\ \HQP\,\Big|\varphi_0 \Big),
\label{Im Veff}
\end{equation}
is very important. 
It is responsible for strong absorption of the low partial-waves, as it will be discussed in detail below. \\

Eq.~(\ref{Im Veff}) can be further reduced by using a spectral expansion of $\HQQ$, 
\[
\HQQ \,\left| q \right\rangle = \varepsilon_q\,\left| q \right\rangle .
\]
One gets
\begin{equation}
W  = -\frac{\pi\,I}{2}\  \sum_{q} \frac{\big(\varphi_0 \big| \mathbb{V}\ \big| q \big\rangle\ \big\langle\tilde{q} \big|\ \mathbb{V}\ \big| \varphi_0 \big)}
{\left( E - \epsilon_{q} \right)^2 + I^{2}/4}.
\end{equation}
The above potential can be approximately evaluated through the following procedures. 
First, the $q$ sum is replaced by an integral over $\epsilon_q$, by introducing the density of states of the CN, 
$\rho_{\scr CN}(\epsilon_q)$  (it should not be confused with the nucleon densities $\rho_{\scr P}({\bf r})$ and 
$\rho_{\scr T}({\bf r})$). 
That is
\begin{equation}
\sum_{q} \rightarrow \int d\epsilon_{q} \ \rho_{\scr CN}(\epsilon_q). 
\end{equation}
The second step is to assume that this density is a slowly varying function of $\epsilon_q$, and take it outside 
the integral. 
Then the $\epsilon_{q}$ integral is just the Lorentzian average of 
$\left( \varphi_0  \right| \mathbb{V} \left| q \right\rangle\, \left\langle \tilde{q} \right| \mathbb{V} \left| \varphi_0 \right\rangle$.
The final approximate expression of the energy-averaged absorptive potential is
\begin{equation}
W= - 2\pi\ \  \overline{ \rho_{q}(\epsilon_q)\ 
\left( \varphi_0  \right| \mathbb{V} \left| q \right\rangle\, \left\langle \tilde{q} \right| \mathbb{V} \left| \varphi_0 \right\rangle
}. 
\label{avpot} 
\end{equation}

It is important to mention that the introduction of energy averaging and the subsequent emergence of a 
complex interaction would seemingly violates flux conservation.
However this is fixed by tracking the path of the lost flux which goes to the formation of the Compound 
nucleus and separately calculate the decay of the latter using the so-called 'statistical theory'. 
The corresponding cross section, the Hauser-Feshbach~\cite{HaF52} cross section contribution to elastic 
scattering (compound elastic) is then added incoherently to the elastic cross section calculated using the 
optical potential equation with the absorptive potential of Eq. (\ref{avpot}). 
This way the flux is accounted for completely.\\

Of course it is a long path to relate the above to $W(r)$ of Eq.(\ref{UW}). 
To begin with, the potential operator of Eq. (\ref{avpot}) is nonlocal when written in configuration space. 
Secondly, it is potentially energy-dependent. 
However, using the above discussion as  a guide, it is customary to use a local approximation for it as 
the imaginary potential of Eq.~(\ref{UW}). 
But the above clearly shows that reference to the compound nucleus formed in the fusion process is important. 
Further, $W(r)$ may account for both closed channels (fusion) absorption and open channels ones (direct reactions). 
Of course these direct channels are accounted for in the Feshbach theory by allowing some of the $Q$-projected 
channels to be open ones. 
We shall not indulge into this procedure here, except to say that the open direct channels would add an additional  
component to the effective interaction, referred to as the {\it dynamic  polarization potential}. 
This component is generally concentrated in the surface region and at the level of $W(r)$ implies a larger diffuseness. 
Accordingly, one has to keep in mind the above observations when using a potential absorption description of fusion.\\

It should be mentioned that the real part of the optical potential is given by the bare potential of Eq.~(\ref{UPP}) 
plus the correction $\Delta U$ (Eq.~(\ref{Delta U})). 
When the couplings are restricted to CN states, this correction is negligible and the imaginary part $W$ is very strong. 
When written in the configuration space, it is, as mentioned before, non-local. 
But a local version is constructed and its range is short. 
It acts exclusively in the inner region of the Coulomb barrier. 
Then, for $l\le l_{\rm cr}$,  the condition for absorption is that the system traverses the barrier of $U_l(r)$. 
On the other hand, the physical interpretation of absorption at partial waves higher than $l_{\rm cr}$ is not clear.\\

The situation is different when couplings with open channels are taken into account. 
Then, the projector $Q$ is split into two terms, one projecting onto closed channels and the other projecting 
onto open ones. 
Following the same procedures as in the case of purely closed channels, one gets and additional
potential, usually called the {\it polarization potential}. 
In case of strong couplings with open channels, as rotational channels in collisions of heavy projectiles
on highly deformed targets, the polarization potential depends strongly on the nuclear structure properties 
of the collision partners. 
Otherwise, the polarization potential has a weak dependence on the collision partners. 
In this case, they may be taken into account by modifying the optical potential. 
As direct reactions take place mainly in grazing collisions, the range of the imaginary potential must be extended. 
In this case, absorption no longer associated exclusively with fusion. 
It gives the total reaction cross section, the sum of fusion with direct reactions.


\subsection{Treatments of absorption}\label{Absorption in potential scattering}



\subsubsection{Absorption by a complex potential}\label{WFWR}


An important issue in quantum mechanical descriptions of scattering is the range of the imaginary potential. 
It depends on the nature of the processes responsible for the attenuation of the incident current in the elastic 
channel, which the imaginary potential simulates. 
In coupled channel calculations including all relevant direct channels, fusion is the only process the imaginary 
potential accounts for. 
In this case, the couplings act at very short distances, in the inner region of the Coulomb barrier. 
Usually, this imaginary potential is represented by the Woods-Saxon function
\begin{equation}
W^{\scr F}(r) = \frac{W_0}{1+\exp\left[ (r-R_0)/a_i \right]},
\label{W(r)-WS}
\end{equation}
with 
\[
R_0 = r_{\scr 0i} \ \left[ A_{\scr P}^{\scr 1/3}\,+\,A_{\scr T}^{\scr 1/3} \right].
\]
The condition of strong absorption with a short range is guaranteed by a large strength parameter, 
say $W_0 = -50$ MeV, and small radius and diffusivity parameters, like $r_{\scr 0i} = 1.0$ fm and 
$a_{\scr i} = 0.20$ fm. \\

On the other hand, in typical optical model analyses, elastic and total reaction cross sections of 
potential scattering calculations are compared with data. 
Of course, the experimental cross sections are influenced by both fusion and direct reactions. 
Thus, the imaginary potential must have a longer range, acting both in the inner region of the 
Coulomb barrier and in the barrier region. 
Then one may use a WS function with larger values of the $r_0$ and $a_0$ parameters, or 
another function with a similar range. 
Alternatively, on can use the same radial dependence of the real potential, and multiply the 
strength parameter by a factor $\lambda$ slightly less than one. 
That is, 
\begin{equation}
W^{\scr R}(r) = i\,\lambda\,U_{\scr N}(r),
\label{W(r)-lambda U}
\end{equation}
where $U(r)$ is the real part of the nuclear interaction and $\lambda$ is a constant, usually 
slightly less then one. 
Gasques {\it et al.}~\cite{GCG06} obtained good descriptions of data  of a large number of 
systems using the above procedure. 
They adopted the S\~ao Paulo potential~\cite{CDH97,CCG02}, with $\lambda = 0.78$.\\


\subsubsection{The IWBC and the WKB approximations}


The other way to account for fusion in potential scattering is to keep the potential real and solve 
the radial equation with an ingoing wave boundary condition. 
The cross sections obtained in this way are very close to the ones obtained with the WKB 
approximation~\cite{TCH17,TCH17a}. 
Both approaches are based on the implicit assumption that fusion is a tunneling phenomenon. 
Thus, one sets
\begin{equation}
\mathcal{P}_{\scr F}(l,E) = \mathcal{T}(l,E),
\label{ps-sigF1}
\end{equation}
where $\mathcal{T}(l,E)$ is the transmission coefficient of a particle with energy $E$ through the 
barrier of the $l$-dependent potential
\begin{equation}
V_l(r) = V_{\scr N}(r) + V_{\scr C}(r) + \frac{\hbar^2}{2\mu\,r^2}\, l(l+1).
\label{Vl}
\end{equation}
For simplicity, instead of Schr\"odinger equations with IWBC, we use Kemble's version of the WKB 
approximation, where the transmission coefficient is given by
\begin{equation}
 \mathcal{T}(l,E) = \frac{1} {1+\exp\left[  2\,\Phi_l(E)\right]},
\label{P-Tl}
\end{equation}
with
\begin{equation}
\Phi_l(E) = \int_{r_1}^{r_2} k_l(r)\ dr.
\label{Phi}
\end{equation}
Above, 
\begin{equation}
k_l(r) = \frac{\sqrt{2\mu\,\left[E - V_l(r) \right]}}{\hbar}
\label{local k}
\end{equation}
is the local wave number and $r_1$ and $r_2$ are respectively the internal and the external turning 
points in the barrier region. 
In addition, there is an innermost turning point, $r_{\rm in}$, located in a region dominated by the 
centrifugal potential. 
The influence of this turning point is disregarded in the IWBC and the WKB descriptions of the collision. 
At energies above the barrier of the potential for the $l^{\rm th}$ partial-wave there are no real values for 
the turning points $r_1$ and $r_2$. 
Then, keeping the integral of Eq.~(\ref{Phi})  along the real-axis one gets $\Phi_l(E) = 0$, and thus 
$\mathcal{T}(l,E) = 1/2$, for any energy above the barrier. 
This problem can be handled with the analytical continuation of the potential on the complex 
$r$-plane~\cite{TCH17,TCH17a}. 
The calculation of the transmission coefficient in this energy range can be simplified if one adopts 
the parabolic approximation for the potential barrier,

\begin{equation}
U_l(r) = B_l -  \frac{1}{2}\, \mu\omega_l^2\ \left( r-R_l \right)^2,
\label{l-parabolae}
\end{equation}
where $B_l $, $R_l$ and $\hbar\omega_l$ are respectively the height, the radius and the curvature 
parameter of the $U_l(r)$ barrier. 
In this case, the transmission coefficient can be evaluated analytically and one finds the so called 
Hill-Wheeler transmission coefficient~\cite{HiW53}
\begin{equation}
T^{\scr HW}(l,E) = \frac{1}{1+\exp\left[ 2\pi\,\left( B_l-E \right)/\hbar\omega_l \right]}.
\label{THW}
\end{equation}
For low partial waves, the potential of Eq.~(\ref{Vl}) has a barrier with maximum $B_l$, located at $R_l$. 
The change of behaviour takes place at the critical angular momentum,  $l_{\rm cr}$, which is the $l$ 
value satisfying the equation
\begin{equation}
\omega_{l} = \sqrt{
-V_l^{\prime\prime}(R_l) /\mu } = 0 .
\end{equation}
The grazing energy for $l_{\rm cr}$, known as the critical energy and denoted by $E_{\rm cr}$, is then given by 
\begin{equation}
E_{\rm cr} = B_{l_{\scr cr}}.
\end{equation}

\bigskip

For  $l > l_{\rm cr}$, the barrier disappears and $\mathcal{T}(l,E)$ vanishes. 
Thus, according to Eq.~(\ref{ps-sigF1}), partial-waves above $l_{\rm cr}$ do not contribute to $\sigma_{\scr F}$. 
Therefore, the partial-wave series of Eq.~(\ref{fusion1}) is truncated at $l = l_{\rm cr}$.  
In this way, $\sigma_{\scr F}(E>E_{\rm cr})$ decreases monotonically with $E$. 
Making the classical approximation for the transmission coefficient, 
$\mathcal{T}\left(l<l_{\rm cr},E>E_{\rm cr} \right) = 1$, the cross section above the critical energy takes the simple form
\begin{equation}
\sigma_{\scr F} \left( E  \right) \simeq \sigma_0\times \frac{E_{\rm cr}}E,
\label{E>Ecr}
\end{equation}
with
\begin{equation}
\sigma_0 = \frac{\pi\hbar^2 \left( l_{\rm cr}+1 \right)^2}{2\mu E}. 
\label{sig0}
\end{equation}
The above expression is very accurate, except for energies just above $E_{\rm cr}$ where the transmission 
coefficients for $l\simeq l_{\rm cr}$ are not yet very close to one.\\

An important difference between fusion probabilities of complex potentials and in the WKB/IWBC approaches 
is that in the former there is an inherent wave reflection from the imaginary part.
This reflection is an effective repulsion which renders the strength to be smaller than the announced one. 


\subsection{The CN formation probability}


According to Eq.~(\ref{Pfus}), absorption by a strong short-range potentials or by IWBC does not guarantees fusion. 
The absorption probability must be multiplied by the CN formation probability, $\mathcal{P}_{\scr CN}$. 
There are two situations where this factor modifies the fusion probability, as discussed below.\\

At near-barrier energies, only low partial-waves contribute to the fusion cross section. 
In this case, absorption probabilities obtained with short-range imaginary potentials and with IWBC
(or equivalently, by the WKB approximation) are very similar. 
The common assumption is that the probability of formation of the CN is unity, once the system overcomes 
the Coulomb barrier. 
However, this depends on the excitation energy of the CN. 
In most systems, however, even at energies around the Coulomb barrier, the CN resonances are strongly 
overlapping and thus the system always find a way to form the CN. 
Accordingly, assuming that the CN formation probability has its maximum value (unity) is quite appropriate. 
Exceptions to this can be found when the CN resonances are isolated (average width of the resonances 
smaller than their average spacing). 
In such a case the CN formation probability, $\mathcal{P}_{\scr CN}(l,E)$, must be less then one. 
This probability depends on the product  $\Gamma_{\scr CN}(E_{CN})\times\rho_{\scr CN}(E_{\scr CN})$, 
where $\rho_{\scr CN}$ is the average density of states of the compound nucleus. 
The average spacing between the resonances is $D_{\scr CN} = 1/\rho_{\scr CN}$. 
The expression for the CN formation probability is roughly given by the Moldauer-Simonius formula~\cite{Mol67,Sim74,Mol75}, 
\begin{equation}
\mathcal{P}_{\scr CN}\left( l,E_{\scr CN}\right) = 1 -e^{-2\pi \alpha},
\label{Mold-Sim}
\end{equation}
with
\begin{equation}
\alpha = \Gamma_{\scr CN}\left( l,E_{\scr CN} \right) \times \rho_{\scr CN}\left( l,E_{\scr CN} \right).
\label{alpha}
\end{equation}
The above indicates that the factor $\mathcal{P}_{\scr CN}\left( l,E_{\scr CN}\right)$ may be very important 
when $\alpha\ll 1$.  \\

\begin{figure}
\begin{center}
\includegraphics*[width=6.5cm]{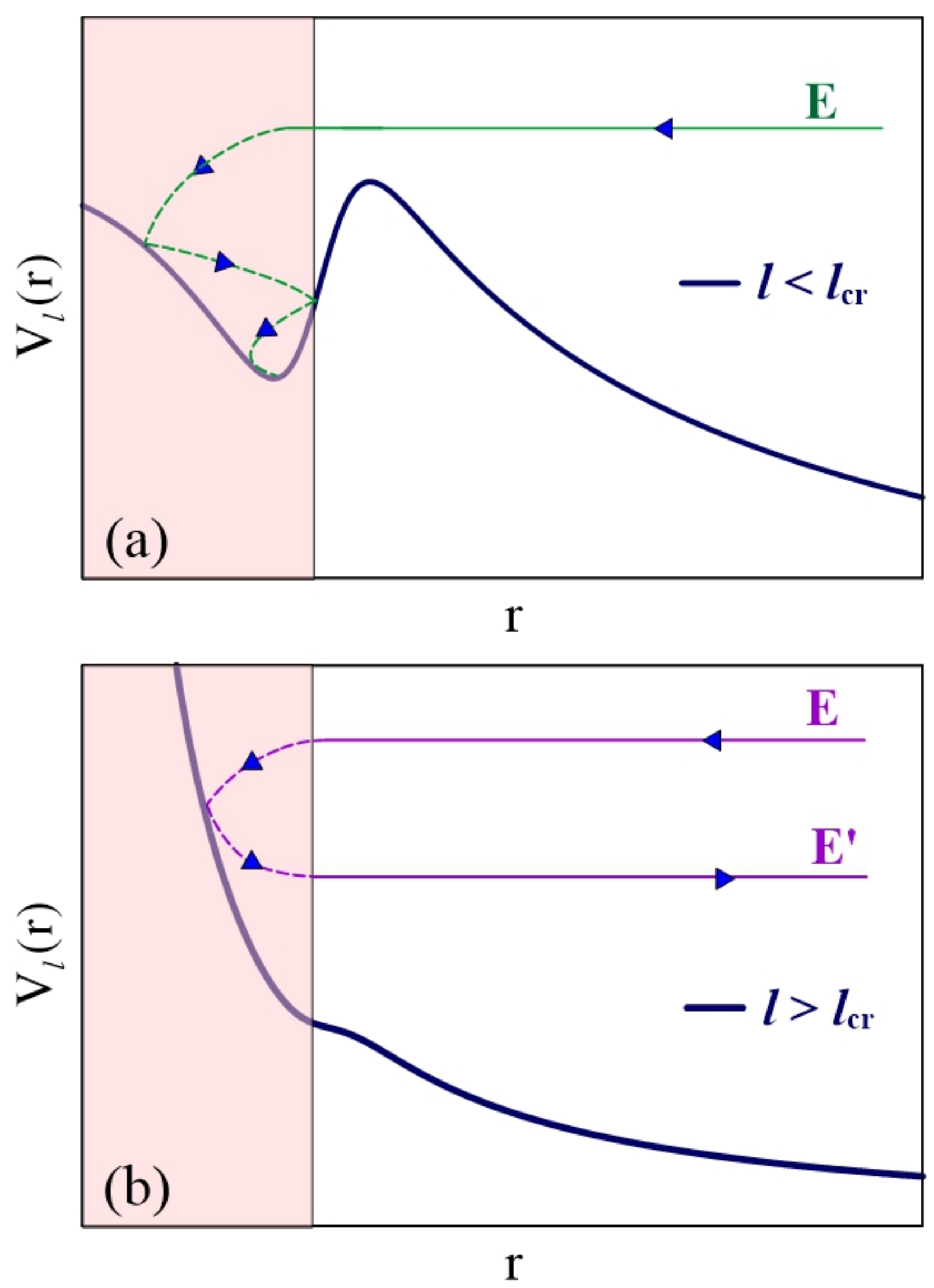}
\end{center}
\caption{(Color on line) Schematic representation of the dissipation of the incident energy. 
Panel (a) shows a collision with $E > B_l$ and $l < l_{\rm cr}$, where it leads to the formation of a
CN. Panel (b) shows a collision with $E > E_{\rm cr}$ and $l > l_{\rm cr}$. 
In this case the system emerges with an energy $E^\prime < E$, without forming a CN.}
\label{CN-no-CN}
\end{figure}
The second situation where the CN formation probability is very important is a collision with 
$E > E_{\rm cr}$ and $l > l_{\rm cr}$. 
For lower energies and partial waves, the kinetic energy of the relative motion is completely 
dissipated when the system reaches the strong absorption region (shaded area in Fig. \ref{CN-no-CN}). 
This energy goes into successive incoherent excitations of single 
particle degrees of freedom of the collision partners (see e.g. Ref.~\cite{BBN78}). 
The system is then caught in the pocked of the potential and, after a long time interval (compared 
with the collision time), the available energy is completely thermalized, forming the CN. 
This situation is depicted in panel (a) of Fig.~\ref{CN-no-CN}. 
In this case, $\mathcal{P}_{\scr CN}(l,E) =1$.

A different process takes place in the collision with energy $E > E_{\rm cr}$ and angular momentum 
higher than $l_{\rm cr}$, represented in panel (b) of Fig.~\ref{CN-no-CN}. 
Owing to the strongly repulsive nature of $U_l(r)$, the stay of the system in the absorption region is 
not long enough for thermalization of the available energy. 
Then, the system looses part of the incident energy and re-separates with an energy $E^\prime < E$. 
In this case there is absorption but no CN formation. 
Thus, $\mathcal{P}_{\scr CN}(l,E) =0$. 
This situation is properly handled in the IWBC and WKB approaches, where partial-waves higher than 
$l_{\rm cr}$ do not contribute to fusion. 
However, it is not correctly described by a strong imaginary potential with a short-range. \\

The hindrance factor given by the Moldauer-Simonious formula (Eq.~(\ref{Mold-Sim})) depends strongly 
on the nuclear structure of the collision partners. 
Further, as mentioned before, in most collisions at near barrier energies this factor is equal to unity. 
On the other hand, the hindrance above the critical angular momentum depends exclusively on the real 
part of the optical potential. 
It can be introduced in calculations with complex potentials through the introduction of a CN formation 
probability given by
\begin{eqnarray}
\mathcal{P}_{\scr CN}(l,E) &=& 1,\ \ {\rm for\ } l \le l_{\rm cr}, \nonumber \\
                                           &=& 0,\ \ {\rm for\ } l > l_{\rm cr} .
\label{PCN_standard}
\end{eqnarray}
We adopt this procedure throughout this paper. 

\bigskip

\begin{figure}
\begin{center}
\includegraphics*[width=8cm]{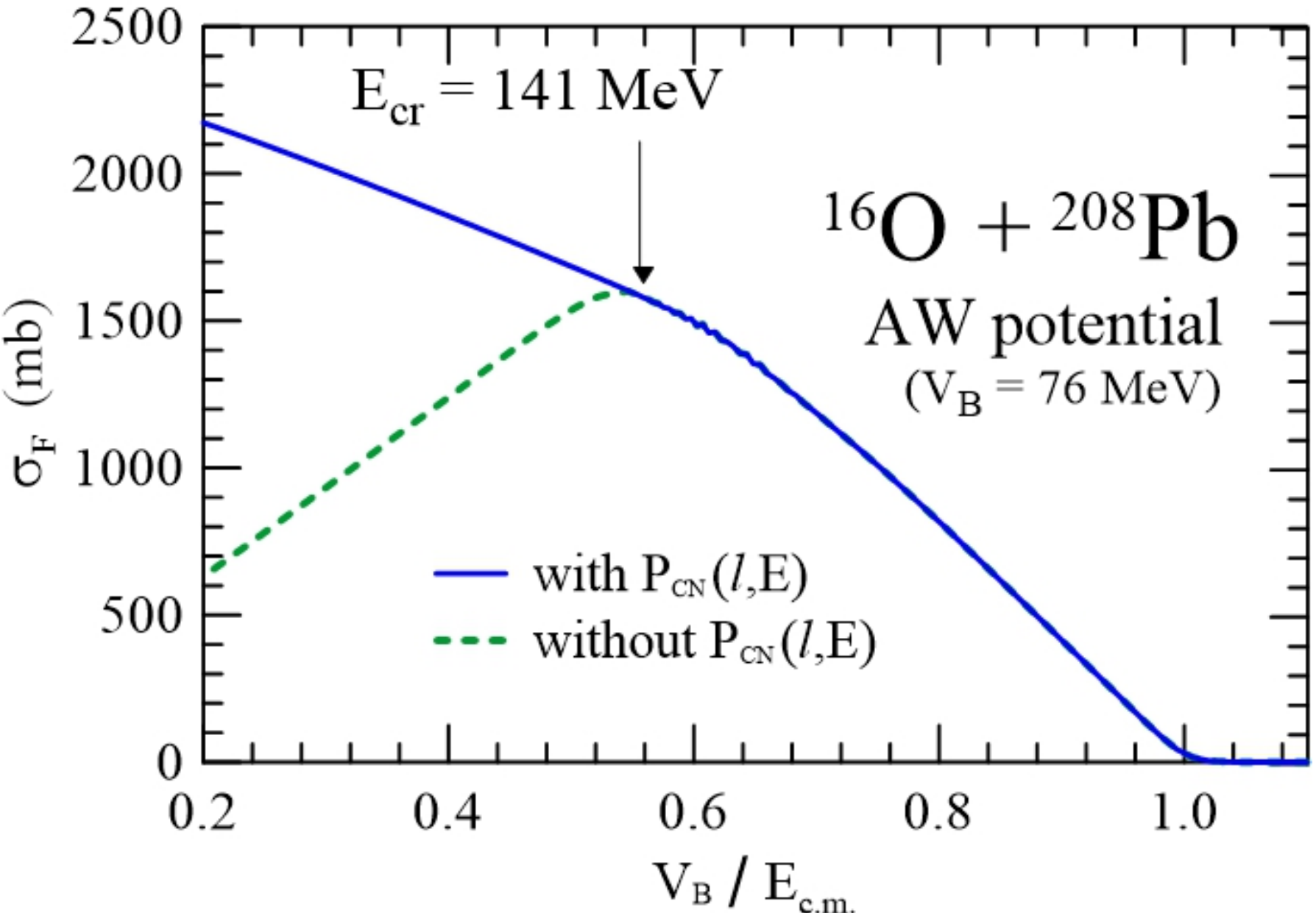}
\end{center}
\caption{(Color on line) Fusion cross sections for the $^{16}$O + $^{208}$Pb system calculated with the 
AW nuclear interaction and the short-range imaginary potential of Eq.~(\ref{W(r)-WS}), with parameters 
$W_0 = -50$ MeV, $r_0 = 1.0$ fm and $a= 0.2$ fm. 
The blue solid line takes into account the CN formation probability of Eq.~(\ref{PCN_standard}) whereas 
the green dashed line does not.}
\label{CN-influence}
\end{figure}
Fig.~(\ref{CN-influence}) illustrates the importance of the CN formation probability in $^{16}$O + $^{208}$Pb fusion. 
The calculations were performed with the AW interaction plus the short-range imaginary potential of Eq.~(\ref{W(r)-WS}). 
For the latter, we adopted the parameters $W_0 = -50$ MeV, $r_0 = 1.0$ fm and $a= 0.2$ fm. 
Usually, fusion and reaction cross sections at energies reaching $E_{\rm cr}$  are plotted against the inverse of the energy. 
We use instead the dimensionless variable $V_{\scr B}/E$. 
As expected, the two curves are indistinguishable at energies below $E_{\rm cr}$, which corresponds to 
$V_{\scr B}/E = 0.54$. 
This corresponds to collisions behaving as in panel (a) of Fig.~\ref{CN-no-CN}. 
However, they are dramatically different above the critical energy and the difference increases as the energy increases. 
This corresponds to the situation of panel (b) of Fig.~\ref{CN-no-CN}. 
Although the absorption increases with energy, the fusion cross section is proportional to $1/E$, following Eq.~(\ref{E>Ecr}).


\section{Application to heavy and light systems}


In this section we investigate the influence of the treatment of absorption and the choice of the nuclear 
potential in fusion cross sections of a heavy ($^{16}{\rm O} + ^{208}{\rm Pb}$) and of a light 
($^{6}{\rm Li} + ^{12}{\rm C}$) system.

\subsection{Dependence of $\sigma_{\scr F}$ on the treatment of absorption}

\begin{figure}
\begin{center}
\includegraphics*[width=8cm]{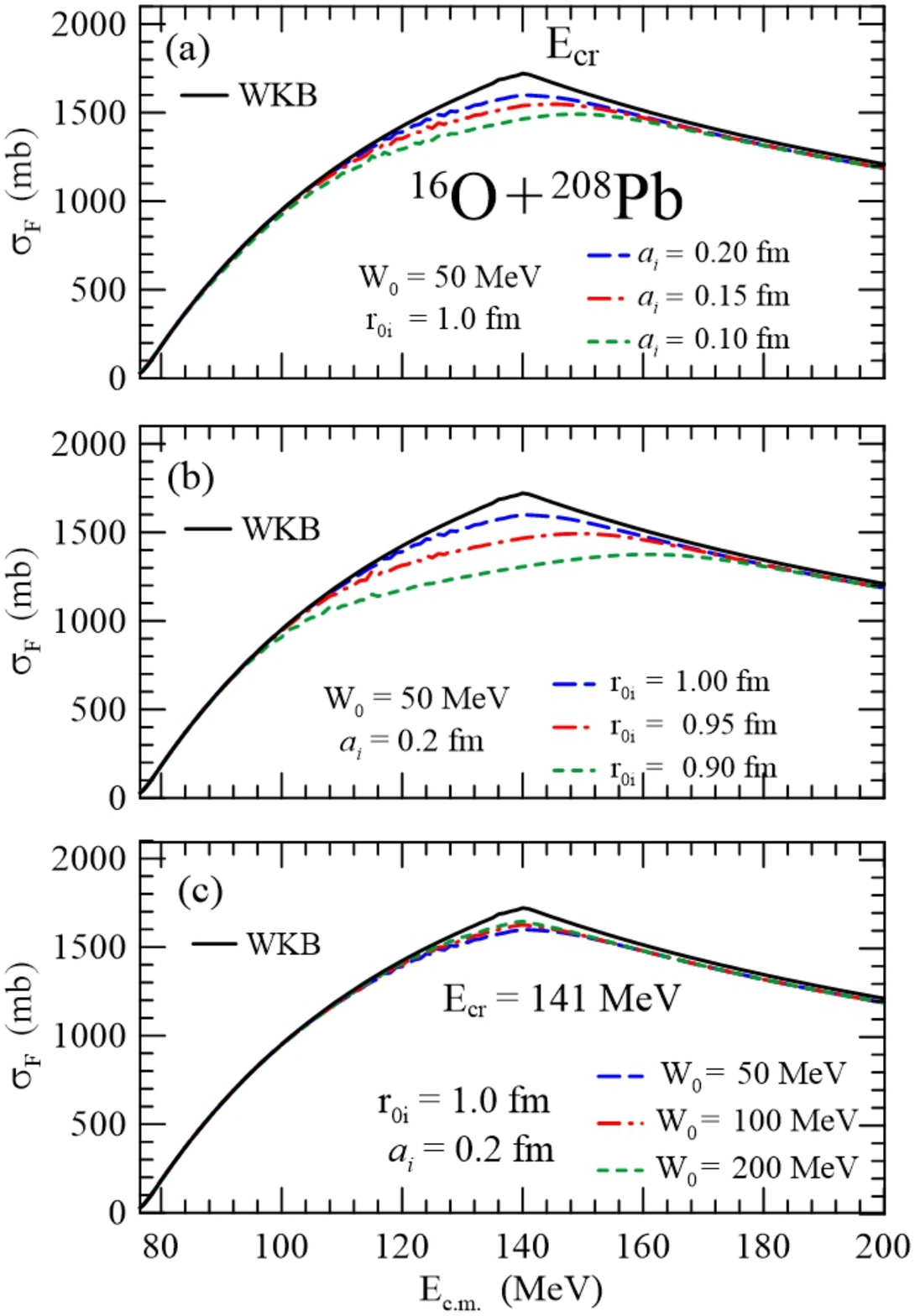}
\end{center}
\caption{(Color on line) Fusion cross sections for the heavy system, $^{16}$O + $^{208}$Pb. 
Panel (a) shows the dependence on the diffuseness parameter, panel (b) the dependence 
on the radius parameter, and panel (c) the dependence on the strength of the imaginary potential. 
The solid black line represents the calculation within the WKB theory. See text for details.}
\label{heavy}
\end{figure}

\begin{figure}
\begin{center}
\includegraphics*[width=8cm]{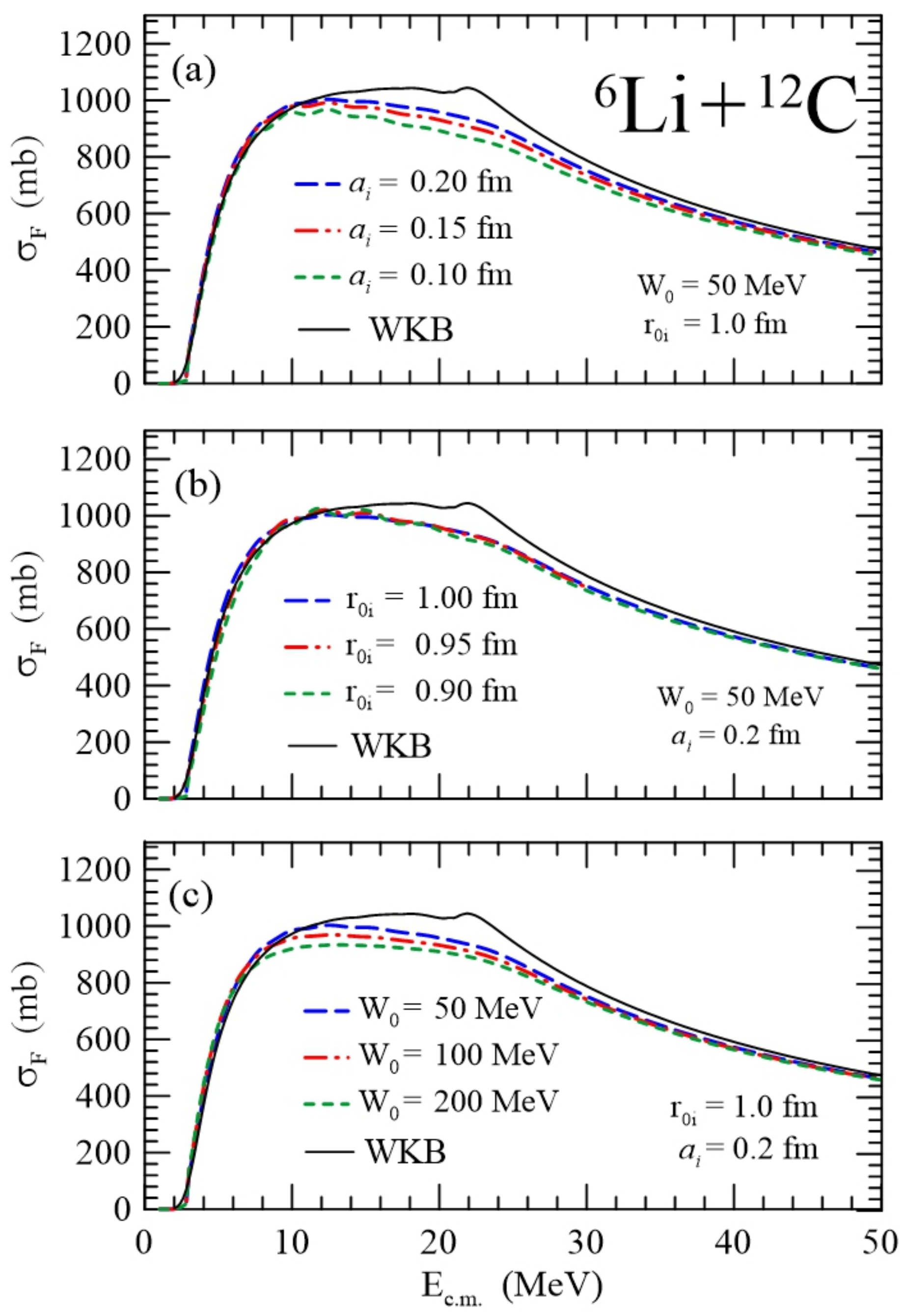}
\end{center}
\caption{(Color on line) Same as Fig.~\ref{heavy} for the light mass system, $^{6}$Li + $^{12}$C. See text for details.}
\label{light}
\end{figure}

Usually, it is assumed that heavy-ion fusion cross sections of barrier penetration models (IWBC or WKB) 
at near-barrier energies are equivalent to cross sections of quantum mechanical calculations with strong 
short-range imaginary potentials, like WS potentials with parameters in the range: 50 MeV $\lesssim W_0\lesssim$ 
200 MeV, 0.9 fm $\lesssim r_{\rm 0i} \lesssim$ 1.0 fm and 0.1 fm $\lesssim a_{\rm i} \lesssim$ 0.2 fm. 
This assumption is not entirely accurate owing to the effect of wave reflection from an absorptive potential. 
In this section, we check this assumption comparing cross sections for one heavy system,  and one light system. 
We use as benchmarks the cross sections of quantum mechanical calculations with WS imaginary potentials 
with the parameters $W_0 = 50$ MeV, $r_{\rm 0i} = 1.0$ fm and $a_{\rm i} = 0.2$ fm. 
In this comparison we adopt the Aky\"uz-Winther potential~\cite{BrW91,AkW81} for the real part of the nuclear interaction. 
With this choice, the Coulomb barriers for the $^{16}{\rm O} + ^{208}{\rm Pb}$ and $^{6}{\rm Li} + ^{12}{\rm C}$ 
systems are $V_{\scr B} = 76.5$ and $V_{\scr B} = 3.3$ MeV, respectively. 
The corresponding critical energies are $E_{\rm cr} = 141$ MeV and $E_{\rm cr} = 21.1$ MeV. 
The fusion cross sections obtained with the different imaginary potentials and with the WKB approximation 
for the two systems are shown in figures \ref{heavy} and \ref{light}.\\

Fig.~\ref{heavy} shows cross sections for the $^{16}$O + $^{208}$Pb system. 
The calculations using the WKB method are shown as solid black lines.
The remaining calculations, using the absorption by a complex potential described
in Section \ref{Absorption in potential scattering}, are represented by dashed lines,
as indicated in the legend.
The benchmark WS results show an agreement with the WKB calculations, up to energies
around 120MeV, followed by a quite different behavior at energies near the critical energy, 
$E_{\rm cr}= 141$ MeV, where the WKB fusion cross section passes abruptly from a steadily 
increasing behavior to a monotonically decreasing one. 
This abrupt change is not observed in the calculations with complex potentials. 
Fusion cross sections derived by both assumptions coincide again for energies above 150 MeV. 
Variations of $a_{\rm i}$, $r_{\rm 0i}$ and  $W_0$, are investigated  in panels (a), (b) and (c), respectively. 
The results show a large sensitivity to the radial parameter, a smaller one to the diffusivity 
and an even smaller one to the strength of the imaginary part of the optical potential.\\

The results of a similar study for the light $^{6}$Li + $^{12}$C system are presented in Fig.~\ref{light}. 
Although the trend of the fusion cross sections calculated within the WKB and absorption methods 
is approximately the same, the relative discrepancies are larger than in the previous case, 
specially at energies around $E_{\rm cr}= 21.1$ MeV. 
We notice that while there is little sensitivity to the radius parameter at all the energies considered, 
there is about the same sensitivity to the diffusivity, and to the strength of  the imaginary potential.\\

We now try to understand these energy dependences. Near the Coulomb barrier, the number of partial 
waves contributing to fusion grows with $E$ and, accordingly, the cross section increases monotonically. 
For a low partial wave, the effective potential has the shape represented on panel (a) of Fig.~\ref{CN-no-CN}.
The potential has a barrier, of height $B_l$, followed by a dip, as $r$ decreases. For an energy $E_0<B_l$, 
fusion is determined by the probability of tunneling trought the barrier. If tunneling occurs, then the fusion 
process takes place, as the system continues its motion drawn by the strong nuclear potential in the well, 
until it reaches the strong absorption region, indicated by the shaded  band at the left of the figure. 
Then the CN is formed.  The outcome is the same at an energy $E > B_l$. In this case the system 
overcomes the barrier and reaches the strong absorption region. There, the kinetic energy is completely
dissipated and the system is caught in the dip of the potential. This situation is schematically represented 
on panel (a) of Fig.~\ref{CN-no-CN}.\\

As the angular momentum increases, the repulsive centrifugal potential leads to a decrease in the depth 
of the effective potential well in the inner region of the Coulomb barrier, until, for $l = l_{cr}$, the potential 
has an inflection point instead of a barrier-well shape. The effective potential for a partial-wave above 
$l_{\rm cr}$ is represented on panel (b) of Fig.~\ref{CN-no-CN}. For a high enough energy, $E$, the system 
overcomes the barrier and reaches the strong absorption region, where its kinetic energy is partly dissipated.
However, there is no CN formation. The system re-separates with an energy $E^\prime$, lower than $E$. 
This process corresponds to the situation schematically represented on panel (b) of Fig.~\ref{CN-no-CN}. 
In this way, partial-waves above $l_{\rm cr}$ do not contribute to fusion. Therefore, the partial-wave series of 
Eq.~(\ref{fusion1}) must be truncated at $l_{\rm cr}$. Then, the cross section at high energies ($E>E_{\rm cr}$) 
decreases linearly with $E$, following Eq.~(\ref{E>Ecr}). This behavior can be observed in Figs. \ref{heavy} 
and \ref{light}. \\

Inspecting Figs. \ref{heavy} and \ref{light}, one concludes that WKB cross sections are very close to the
ones obtained by quantum mechanics with strong absorption potentials, at near-barrier energies and at 
energies well above $E_{\rm cr}$. However, they differ at energies in the neighbourhood of $E_{\rm cr}$. 
Further, the quantum mechanical cross sections in this region depend on the parameters of the imaginary
potential. The discrepancies in this energy range arise from reflections by the real and the imaginary parts 
of the potentials at partial waves just below $l_{\rm cr}$. In this 
case there is a single turning point very close to the barrier radius, outside the region of strong absorption.
Then the incident wave is reflected without contributing to fusion. WKB calculations neglect this effect
and for this reason it overestimates the fusion cross section in the neighbourhood of $E_{\rm cr}$. At
higher energies the situation is different. The turning point for that same partial wave moves into the strong 
absorption region, and then the fusion probability is close to one both in the WKB and in the quantum 
mechanical calculations. 

\subsection{Dependence of $\sigma_{\scr F}$ on the nuclear potential}\label{AWvsSPP}

\begin{figure}
\begin{center}
\includegraphics*[width = 8.5 cm]{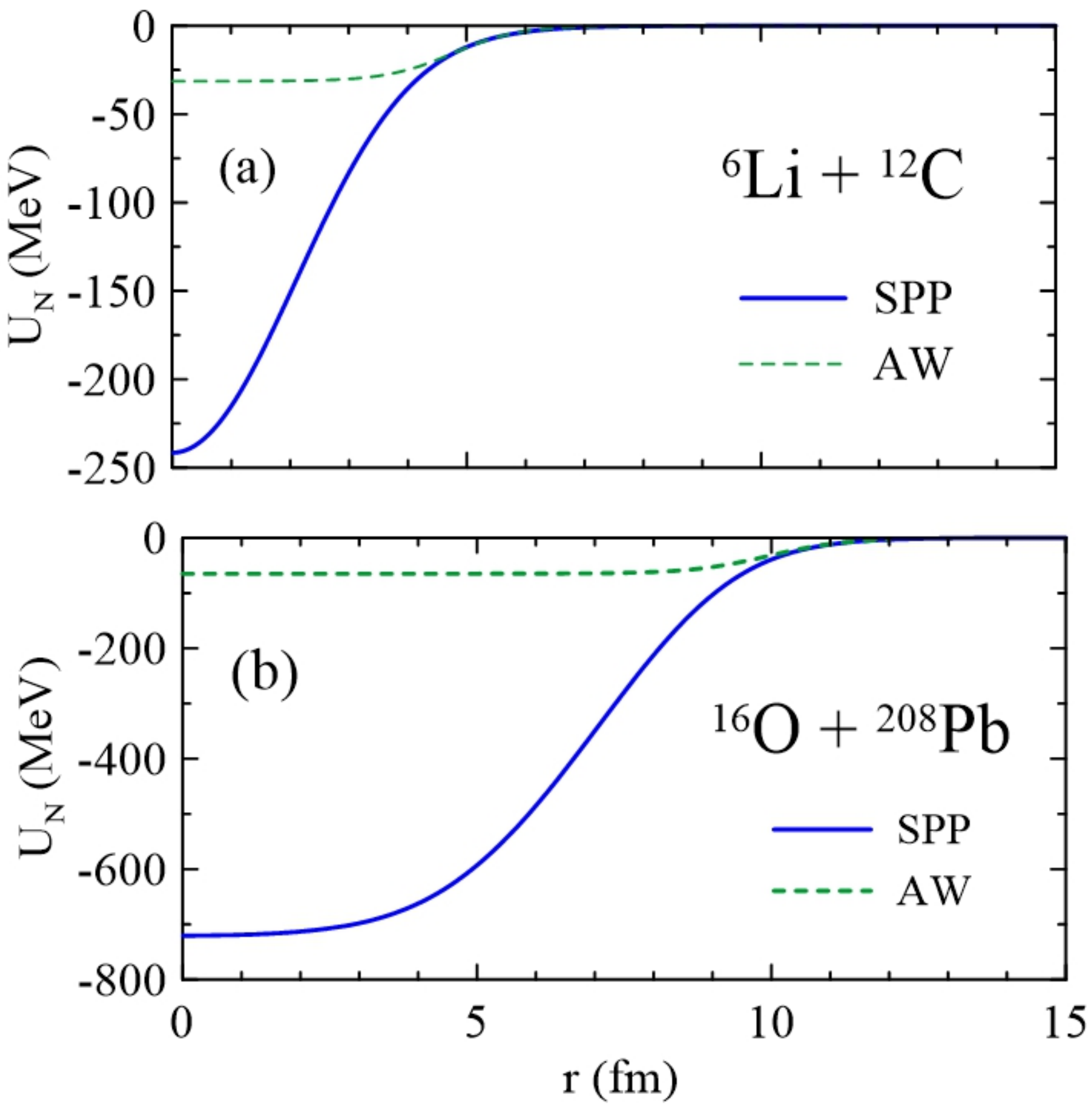}
\end{center}
\caption{(Color on line) The AW and the SPP potentials for the $^{6}$Li + $^{12}$C and $^{16}$O + $^{208}$Pb systems.}
\label{AW-vs-SPP}
\end{figure}

Now we investigate the sensitivity of the fusion cross sections to the choice of the nuclear potential. 
We study the $^{16}$O + $^{208}$Pb and $^6$Li+$^{12}$C systems,
performing calculations with complex potentials. We consider the AW and the SPP nuclear 
interactions, and adopt the short-range imaginary potential of Eq.~(\ref{W(r)-WS}),
with the parameters: $W_0 = 50$ MeV, $r_0 = 1.0$ fm and $a = 0.2$ fm.\\
Fig. \ref{AW-vs-SPP} illustrates the situation for these two bare nuclear potentials
that coincide on the external region, but that have very different predictions for the inner well. 
\\

Table  \ref{Tab1} shows the heights, radii and curvature parameters 
in the parabolic fits of the AW and SPP barriers. It shows also the critical energy in each case. 
As we have noticed, the barrier parameters for the two potentials are quite similar, for both systems. 
Further, the critical energies predicted by the two potentials for the light $^6$Li+$^{12}$C 
system are not very different. The one predicted by the SPP is a roughly 20\% higher. 
However, the critical energies for the heavy  $^{16}$O + $^{208}$Pb system are, indeed, very 
different, with the prediction of the SPP being almost three times larger than the value predicted 
by the AW potential. 
The reason why the critical energies for the SPP are systematically higher is that this potential is 
much deeper than the AW. 
Whereas the depth of the former is of a few tens of MeV, that for the latter is a few hundreds of MeV. 
This difference becomes progressively more important as the system mass grows, as illustrated in  
Fig.~\ref{AW-vs-SPP}.
The critical energy determines the transition between two different energy regimes of the fusion 
cross section, referred to as region 1 (near barrier) and region 2 (above $E_{\scr cr}$). 
Since this transition can be observed in the data, the experimental determination of the transition 
energy can be used as a criterium to select appropriate models for the bare potential. \\
\begin{table}
\caption{The parameters of the parabolic approximation for the Coulomb barriers of the 
systems considered in the present work. The corresponding critical energies are also shown.
}
\centering
\begin{tabular} [c] {lcc}
\hline 
System:                                               &          $\qquad ^{16}$O + $^{208}$Pb\ \ \ \ \ \             &   $\qquad^{6}$Li + $^{12}$C\ \ \ \   \\ 
 \hline 
    $V_{\scr B}$ (MeV) $\qquad$ \ \     &                                                                         &                                                 \\
      $\qquad$           AW:               \ \    &           76.5                                                       &         3.3                               \\
      $\qquad$           SPP:            \  \    &           76.0                                                      &          3.0                                \\
 \hline
    $R_{\scr B}$ (fm)     \ \ \ \    \           &                                                                          &                                                 \\
      $\qquad$           AW:            \ \      &              11.6                                                      &          7.4                                 \\
      $\qquad$           SPP:         \  \      &              11.7                                                      &          7.7                                   \\
 \hline
    $\hbar\omega$ (MeV)                   &                                                                          &                                                    \\
      $\qquad$           AW:            \ \    &                4.5                                                   &            2.7                                \\
      $\qquad$           SPP:       \  \      &                 4.6                                                   &           2.9                                 \\
 \hline
    $E_{\scr cr}$ (MeV)      \ \ \ \        &                                                                          &                                                       \\
      $\qquad$           AW:           \ \    &                 141                                                   &         21.1                                       \\
      $\qquad$           SPP:   \ \          &                  384                                                 &          26.3                                      \\
    \hline 
\end{tabular}
\label{Tab1}
\end{table}

\begin{figure}
\begin{center}
\includegraphics*[width=9cm]{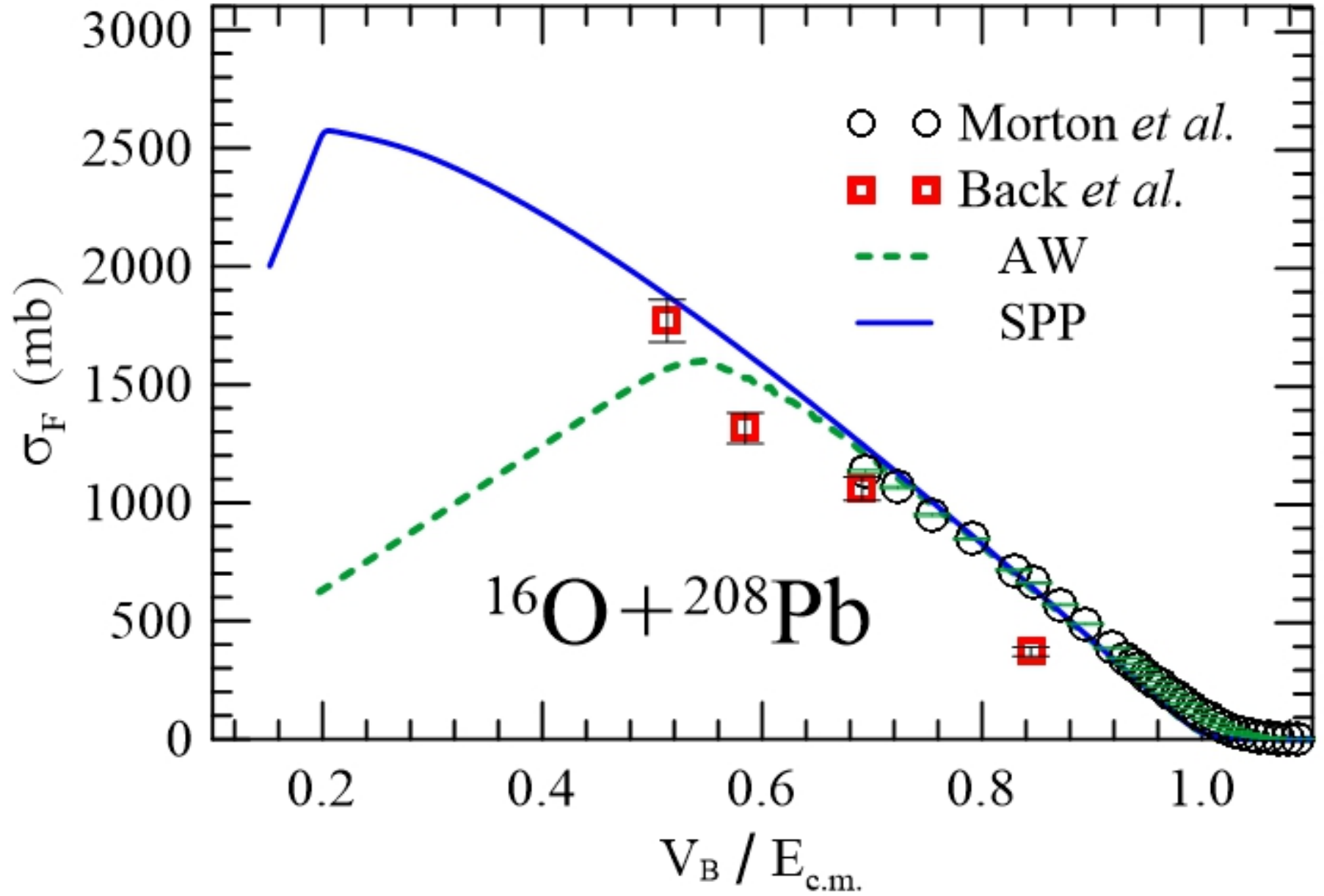}
\end{center}
\caption{(Color on line) Theoretical fusion cross sections for the $^{16}$O + $^{208}$Pb system, 
calculated with the AW (green dashed line) and the SPP (blue solid line). 
The experimental data of Morton {\it et al.}~\cite{MBD99} (black open circles) and 
Back {\it et al.}~\cite{BBG85} (red squares) are also shown.}
\label{data1}
\end{figure}

Having this in mind, we compared the cross sections calculated with the AW and the SPP potentials 
with the available data.  
Fig.~\ref{data1} shows the comparison for the $^{16}$O + $^{208}$Pb system. 
The figure shows the AW cross section (green dashed line) in comparison with the SPP cross section 
(blue solid line) and the data of Morton {\it et al.}~\cite{MBD99} (black open circles), and of 
Back {\it et al.}~\cite{BBG85} (red squares). 
The data of Ref.~\cite{MBD99} is restricted to the low-energy region, where the two theoretical curves 
are very close, and they agree very well with the theory. 
The older data of Ref.~\cite{BBG85} reaches higher energies. 
Three data points were taken at energies below the critical energy for the two potentials and they are 
a little lower than the theoretical curves. 
The fourth data point was taken at an energy slightly higher that the critical energy of the AW potential, 
and it seems to follow the growing trend of the SPP cross section. 
However, considering that there is a single point in this region, the comparison of the data with the 
theoretical curves is not conclusive.

\begin{figure}
\begin{center}
\includegraphics*[width=9cm]{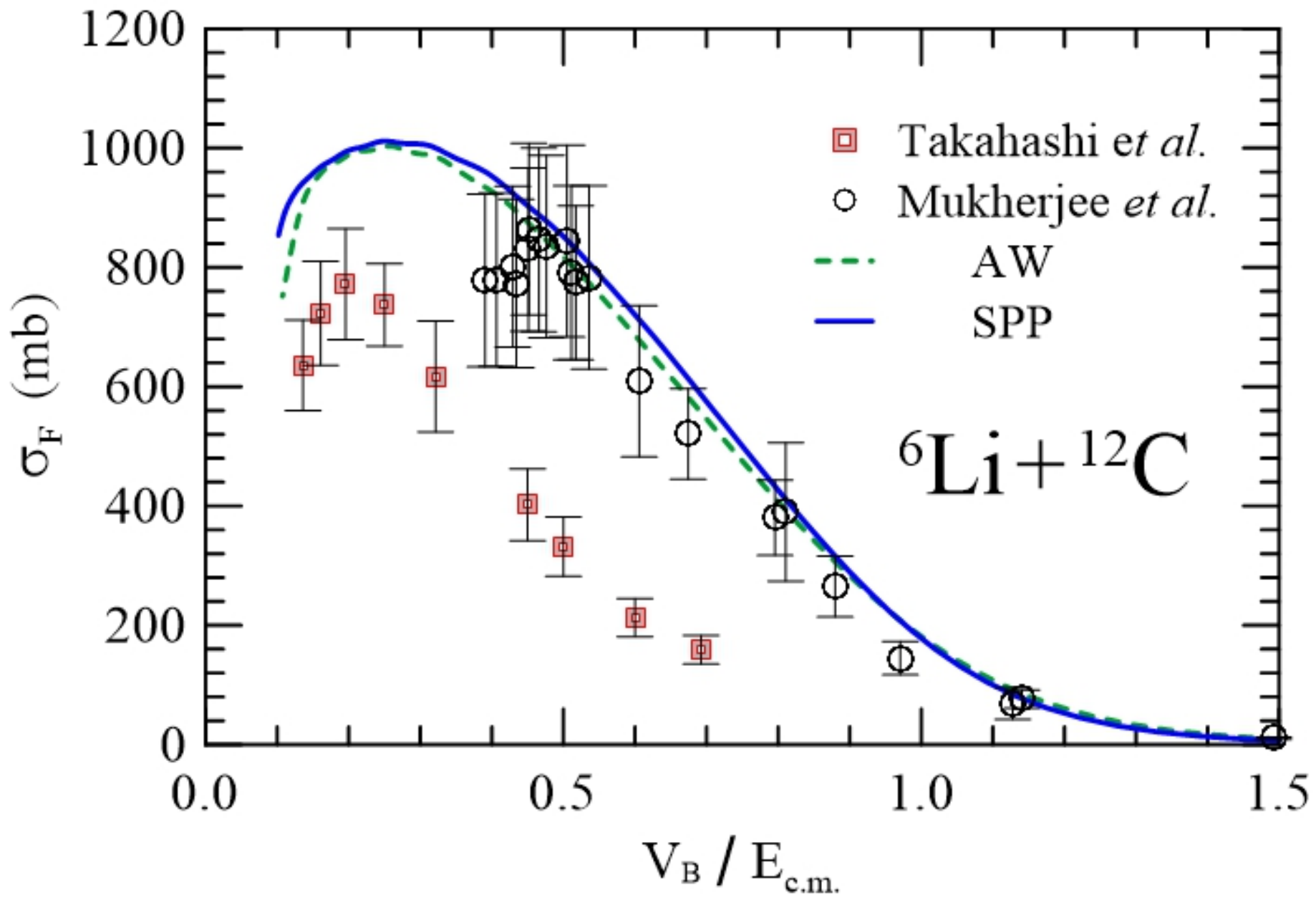}
\end{center}
\caption{(Color on line) Theoretical fusion cross sections for the $^6$Li + $^{12}$C system compared 
with the available data. 
The notation of the curves is the same as in the previous figure and now the black open circles and the 
red square correspond respectively to the data of  of Mukherjee {\it et al.}~\cite{MPC98} and of 
Takahashi  {\it et al.}~\cite{TMS97}. }
\label{data2}
\end{figure}

\bigskip

Fig.~\ref{data2} shows an analogous comparison of cross sections for the $^{6}{\rm Li} + ^{12}{\rm C}$ system. 
The notation of the two curves is the same, whereas the black open circles and  the red squares correspond 
respectively to the data of Mukherjee {\it et al.}~\cite{MPC98} and Takahashi~{\it et al.}~\cite{TMS97}. 
First, one notices that the two curves are very close in the whole energy range. 
This is consistent with the similar barriers of the two potentials.  
The transition energies of the two potentials are also close. 
The data of Mukherjee {\it et al.} agree very well with the theoretical predictions of the two potentials. 
However, the data points at the highest energies seem to indicate a transition to region 2 before the 
theoretical predictions.
Nevertheless they are still in agreement with the theoretical curves within the error bars. 
The data of Takahashi {\it et al.} are systematically lower than the theoretical curves, and also then
the data of Mukherjee {\it et al.}. 
On the other hand, they indicate a transition energy consistent with  theoretical predictions.

\section{Discussion and Conclusions}

This paper presents a detailed investigation of the sensitivity of fusion cross sections to different commonly 
used nuclear potentials and treatments of absorption in potential scattering. 
We evaluate fusion cross sections for the light $^6$Li + $^{12}$C system and for the heavier system 
$^{16}$O + $^{208}$Pb. 
We performed calculations at energies ranging from the Coulomb barrier to beyond the critical energy, 
above which the effective potential for the grazing angular momentum ceases to have a pocket.\\

We compare cross sections of the WKB approximation with those obtained with quantum mechanical 
calculations with complex potentials. 
We point  out that the partial-wave series giving the fusion cross section must be truncated at the critical 
angular momentum, since contribution from higher partial waves have no physical meaning. 
We find that WKB cross sections are similar to the ones obtained from quantum mechanical
calculations with different short-range strong absorption potentials, except in a neighborhood 
of the critical energy. 
In this region,  the WKB cross sections are systematically higher than the quantum mechanical calculations,
which  show some dependence on the parameters of the imaginary potential.\\

We performed quantum mechanical calculations of fusion cross sections using different versions of the 
folding interaction: the Aky\"uz-Winther and the S\~ao Paulo potentials. 
The calculated cross sections where compared to each other and to  the available experimental data. 
At near-barrier energies, the theoretical cross sections for the two systems are very close, and also 
close to most data. 
At higher energies, the situation for the $^6$Li + $^{12}$C system did not change. 
However, the theoretical cross sections for $^{16}$O + $^{208}$Pb where rather different. 
The AW cross section starts to decrease at $E\sim141$ MeV ($V_{\scr B}/E \sim 0.54$) whereas the 
one associated to the SPP keeps growing until $E=384$ MeV  ($V_{\scr B}/E \sim 0.20$). 
The difference is a consequence of the deeper well in the SPP, which leads to a higher critical energy. 
An important consequence is that one can obtain invaluable information about the nuclear interaction 
at small distances comparing theoretical predictions with the data. 
Unfortunately, the presently available data for the $^{16}$O + $^{208}$Pb system are restricted to the 
relatively low energy region, where the two theoretical cross section are close. 
Therefore, experiments at higher energies should be important to help understand the nuclear potential
between heavy ions at short separations.

\vskip 1cm


\section*{Acknowledgments}
Work supported in part by the CNPq, CAPES, FAPERJ, FAPESP (Brazil), and by the PEDECIBA and ANII (Uruguay). The Brazilian authors 
acknowledge also partial financial support from the INCT-FNA (Instituto Nacional de Ci\^encia e Tecnologia- F\'\i sica Nuclear e Aplica\c c\~oes), 
research project 464898/2014-5. M. S. H. acknowledges support from the CAPES/ITA Senior Visiting Professor Fellowship Program. 

\bibliographystyle{apsrev}

\end{document}